\documentclass[12pt]{article}

\begin{document}

%%%%%%%%%%%%%%%%%%%%%%%%%%
% Vatche Sahakian's macros

\newcommand{\bb}{\begin{equation}}
\newcommand{\ee}{\end{equation}}
\newcommand{\bbb}{\begin{eqnarray}}
\newcommand{\eee}{\end{eqnarray}}
\newcommand{\diag}{\mbox{diag }}
\newcommand{\Str}{\mbox{STr }}
\newcommand{\Tr}{\mbox{Tr }}
\newcommand{\Det}{\mbox{Det }}
\newcommand{\C}[2]{{\lk [{#1},{#2}\re ]}}
\newcommand{\AC}[2]{{\lk \{{#1},{#2}\re \}}}
\newcommand{\kk}{\hspace{.5em}}
\newcommand{\vc}[1]{\mbox{$\vec{{\bf #1}}$}}
\newcommand{\mc}[1]{\mathcal{#1}}
\newcommand{\del}{\partial}
\newcommand{\lk}{\left}
\newcommand{\ave}[1]{\mbox{$\langle{#1}\rangle$}}
\newcommand{\re}{\right}
\newcommand{\pd}[1]{\frac{\del}{\del #1}}
\newcommand{\pdd}[2]{\frac{\del^2}{\del #1 \del #2}}
\newcommand{\Dd}[1]{\frac{d}{d #1}}
\newcommand{\sech}{\mbox{sech}}
\newcommand{\pref}[1]{(\ref{#1})}

\newcommand
{\sect}[1]{\vspace{20pt}{\LARGE}\noindent
{\bf #1:}}
\newcommand
{\subsect}[1]{\vspace{20pt}\hspace*{10pt}{\Large{$\bullet$}}\mbox{ }
{\bf #1}}
\newcommand
{\subsubsect}[1]{\hspace*{20pt}{\large{$\bullet$}}\mbox{ }
{\bf #1}}

\def\ie{{\it i.e.}}
\def\eg{{\it e.g.}}
\def\cf{{\it c.f.}}
\def\etal{{\it et.al.}}
\def\etc{{\it etc.}}

\def\e{{\mbox{{\bf e}}}}
\def\AA{{\cal A}}
\def\BB{{\cal B}}
\def\CC{{\cal C}}
\def\DD{{\cal D}}
\def\EE{{\cal E}}
\def\FF{{\cal F}}
\def\GG{{\cal G}}
\def\HH{{\cal H}}
\def\II{{\cal I}}
\def\JJ{{\cal J}}
\def\KK{{\cal K}}
\def\LL{{\cal L}}
\def\MM{{\cal M}}
\def\NN{{\cal N}}
\def\OO{{\cal O}}
\def\PP{{\cal P}}
\def\QQ{{\cal Q}}
\def\RR{{\cal R}}
\def\SS{{\cal S}}
\def\TT{{\cal T}}
\def\UU{{\cal U}}
\def\VV{{\cal V}}
\def\WW{{\cal W}}
\def\XX{{\cal X}}
\def\YY{{\cal Y}}
\def\ZZ{{\cal Z}}

\def\sinh{{\rm sinh}}
\def\cosh{{\rm cosh}}
\def\tanh{{\rm tanh}}
\def\sgn{{\rm sgn}}
\def\det{{\rm det}}
\def\trace{{\rm Tr}}
\def\exp{{\rm exp}}
\def\sh{{\rm sh}}
\def\ch{{\rm ch}}

\def\ell{{\it l}}
\def\str{{\it str}}
\def\lp{\ell_{{\rm pl}}}
\def\blp{\overline{\ell}_{{\rm pl}}}
\def\ls{\ell_{{\str}}}
\def\bls{{\bar\ell}_{{\str}}}
\def\bM{{\overline{\rm M}}}
\def\gs{g_\str}
\def\gym{{g_{Y}}}
\def\geff{g_{\rm eff}}
\def\eff{{\rm eff}}
\def\r11{R_{11}}
\def\kel{{2\kappa_{11}^2}}
\def\kten{{2\kappa_{10}^2}}
\def\lpten{{\lp^{(10)}}}
\def\alp{{\alpha '}}
\def\alpe{{{\alpha_e}}}
\def\le{{{l}_e}}
\def\aleff{{\alp_{eff}}}
\def\sqaleff{{\alp_{eff}^2}}
\def\tgs{{\tilde{g}_s}}
\def\talp{{{\tilde{\alpha}}'}}
\def\tlp{{\tilde{\ell}_{{\rm pl}}}}
\def\tr11{{\tilde{R}_{11}}}
\def\wtilde{\widetilde}
\def\what{\widehat}
\def\hlp{{\hat{\ell}_{{\rm pl}}}}
\def\hr11{{\hat{R}_{11}}}
\def\hf{{\textstyle\frac12}}
\def\coeff#1#2{{\textstyle{#1\over#2}}}
\def\CY{Calabi-Yau}
\def\lessapprox{\;{\buildrel{<}\over{\scriptstyle\sim}}\;}
\def\greaterapprox{\;{\buildrel{>}\over{\scriptstyle\sim}}\;}
\def\inbar{\,\vrule height1.5ex width.4pt depth0pt}
\def\IC{\relax\hbox{$\inbar\kern-.3em{\rm C}$}}
\def\IR{\relax{\rm I\kern-.18em R}}
\def\IP{\relax{\rm I\kern-.18em P}}
\def\Z{{\bf Z}}
\def\R{{\bf R}}
\def\One{{1\hskip -3pt {\rm l}}}
\def\sst{\scriptscriptstyle}
\def\osc{{\rm\sst osc}}
\def\lam{\lambda}
\def\lc{{\sst LC}}
\def\pr{{\sst \rm pr}}
\def\cl{{\sst \rm cl}}
\def\D{{\sst D}}
\def\bh{{\sst BH}}
\def\vev#1{\langle#1\rangle}

\begin{titlepage}
\rightline{CLNS 01/1771}

\rightline{hep-th/0112063}

\vskip 2cm
\begin{center}
\Large{{\bf Strings in Ramond-Ramond backgrounds
}}
\end{center}

\vskip 2cm
\begin{center}
Vatche Sahakian\footnote{\texttt{vvs@mail.lns.cornell.edu}}
\end{center}
\vskip 12pt
\centerline{\sl Laboratory of Nuclear Studies}
\centerline{\sl Cornell University}
\centerline{\sl Ithaca, NY 14853, USA}

\vskip 2cm

\begin{abstract}
We write the type IIB worldsheet action
in classes of
bosonic curved backgrounds threaded with Ramond-Ramond fluxes. 
The fixing of the kappa symmetry in the light-cone
gauge and the use of the Bianchi identities
of the supergravity theory lead to an expression of a
relatively simple form, yet rich with new physical information about how
fundamental strings react to the presence of RR fields. The results
are useful in particular to the study of vacuum structure and dynamics
in the context of the Holographic duality; and to possibly formulate an
open-closed string duality at the level of the worldsheet.

\end{abstract}

\end{titlepage}
\newpage
\setcounter{page}{1}

\section{Introduction and Results}
\label{intro}

\vspace{0.5in}
\begin{center}
NOTE ADDED
\end{center}
The reader is referred to hep-th/0402037 for complete results, including the full form of action
and comparison with literature.

\vspace{0.5in}
In many realizations of 
the Holographic duality~\cite{MALDA1,KLEB,WITHOLO}, 
where a perturbative
string theory is found dual to strongly coupled dynamics in a field theory
or in 
another string theory, the closed strings on the weakly coupled
side of the duality are emersed in background Ramond-Ramond (RR) fluxes.
Knowledge of the couplings of strings to such fluxes is then an important 
ingredient in exploring the principles of the duality itself
and in understanding the dual theory. 

The matter becomes particularly urgent with the discovery of Non-Commutative
Open String (NCOS) theories~\cite{GMSS,KLEBMALDA,SST}. 
In these settings, a perturbative definition
of a new theory of open strings is known -- it being inherited from the parent 
string theory whose particular low energy truncation the NCOS dynamics 
corresponds to; and a definition of this open string
theory at strong coupling is
provided via a perturbative closed string theory in a certain curved
background with RR fluxes~\cite{HARMARKS,VVSNCOS}. 
Information about the particular forms of the couplings
to these fluxes in the closed string sigma model 
holds the promise to help us
better understand strong coupling dynamics of an open string theory,
and to correspondingly formulate the open-closed string duality as a map
at the level of the worldsheet.

There are three main
approaches in writing down an action of closed superstrings in
an arbitrary background. In the RNS formalism, powerful 
computational techniques
are available, yet the vertex operators sourced by RR fields involve
spin fields. Consequently, the resulting action is not terribly useful
in practice. A second approach is the GS formalism with spacetime
supersymmetry, generally leading
to actions that are useful in unraveling the
semi-classical dynamics of the sigma model. On the down side, manifest
Lorentz symmetry is lost with the choice of the light-cone gauge; and,
at one loop level for example, the lost symmetry results in
serious complications. The third approach was developed 
recently~\cite{BVW}
and involves a hybrid picture; in this strategy, part of the spacetime
symmetries remain manifest yet couplings to the RR fields take 
relatively simple forms.
The cost is the introduction of several auxiliary fields, 
and certain assumptions on the form of the background.

In this work, we focus on the second GS method with spacetime supersymmetry
and on determining the component form of the action. Our interest is to
eventually study, semi-classically, closed string dynamics in certain
backgrounds that may not be endowed with a lot of symmetries. Other
attempts involve capitalizing on the large amount of symmetry present, in
particular, in AdS spaces (see, for example, \cite{RAHMRAJ}-\cite{PESANDO}).
Most of the difficulties involved in writing down this string action
in general form are due to the fact that superspace
in the presence of supergravity, while still being an attractive
setting, can be
considerably elaborate~\cite{SUSYBOOK}. 
A large amount of superfluous symmetries need to
be fixed and computations are often prohibitavily lengthy.

The task is significantly simplified by the use of the method
of normal coordinate
expansion~\cite{MUKHINLSM,GRISARU} 
in superspace. This was developed for the Heterotic
string in~\cite{ATICKDHAR}, and, along with the use of computers for analytical
manipulations, makes it straightforward to determine the type IIA and IIB 
sigma models as well. The additional complications
that arise in these cases, and that are absent in the Heterotic string case,
are due entirely to the presence of the RR fields. 
We concentrate for now on the IIB theory. In~\cite{CVETIC}, 
part of this action, to quadratic order in the fermions, was derived
starting from the supermembrane action and using T-duality. With
different methods and starting from
superspace in IIB theory directly, we will write the full-form of this action
in the light-cone gauge relevant to most backgrounds of interest.

There are two additional steps within
the normal coordinate expansion technique that help to simplify matters
further.
One involves fixing the $\kappa$ symmetry early on in the computation.
We will show that this truncates the action to quartic order in the
fermions. The second step involves making certain general assumptions on the
form of the background fields. These assumptions are crucial; otherwise,
expressions explodes in size by many orders of magnitude. 

The form of the background fields we focus on are inspired by~\cite{ATICKDHAR} 
and by the need to apply the results to settings that arise in
the context of the Holographic duality. In particular, 
fields generated by electric and magnetic
D-branes of various configurations share certain general features of interest.
We list all the conditions we require on the background fields so that
the results in this work are applicable:
\begin{itemize}
\item The fermionic fields must be zero. In particular, 
the gaugino and gravitino of the IIB theory have no condensates.

\item We choose a certain space direction that, along with the time
coordinate, we will associate with the light-cone gauge fixing later.
We refer to the other eight spatial directions as being transverse to the
light-cone. Then all background fields must depend only on the transverse
coordinates.

\item Tensorial fields can have indices in the transverse directions;
and in the two light-cone directions only if the light-cone coordinates
appear in pairs.

\item We assume that the metric can be put into diagonal form.
\end{itemize}

If we were to consider, for example, a background 
consisting of a number of Dp branes,
we choose the light-cone directions {\em parallel} to the worldvolume
of the branes. All conditions listed above are then satisfied. 
The conditions are of course satisfied in
more general cases than this particular example.

Under these assumptions, and once the $\kappa$ symmetry is fixed,
the IIB action takes the form
\bb\label{genform}
I=I^{(0)}+I^{(2)}+I^{(4)}+J^{(4)}\ ,
\ee
where the superscripts denote the number of fermionic fields in each part.
The first term is the standard bosonic part\footnote{Note that
the signature of the metric we use is `unconventional'. See Appendix
A for details.}
\bb\label{I0}
I^{(0)}=\int d^2\sigma\ \lk[ 
\frac{1}{2} \sqrt{-h} h^{ij} G_{mn} \del_i x^m \del_j x^n
+\frac{1}{2} \varepsilon^{ij} \del_i x^m \del_j x^n b^{(1)}_{nm}\re]\ .
\ee
We represent the two spacetime spinors by a single Weyl -- but otherwise
complex -- 16 component spinor $\theta$.
At quadratic order in $\theta$, the action involves the following
couplings to the background fields
\bbb
I^{(2)}&=&\int d^2\sigma\ \lk[ -i V^+_{i} 
\lk(
\varepsilon^{ij} \theta \sigma^- \hat{D}_j \theta
+\sqrt{-h} h^{ij} \bar{\theta} \sigma^- \hat{D}_j \theta\re)\re.\nonumber\\
&+&\omega^2 V^+_i V^a_j\times \lk( \re.
- \sqrt{-h} h^{ij}Z_{abcde}
\bar{\theta} \sigma^- \sigma^{bcde}\theta\nonumber \\
&+&8 \sqrt{-h} h^{ij}Z^{-+}_{\hspace{15pt}bcd}
\bar{\theta} \sigma^- \sigma_a^{\hspace{8pt}bcd}\theta
-48  \varepsilon^{ij}Z^{-+}_{\hspace{15pt}a b c}
\theta \sigma^- \sigma^{bc}\theta\nonumber \\
&+&i \varepsilon^{ij} \bar{f}^{-+}_{\hspace{15pt}b}
\bar{\theta} \sigma^- \sigma_a^{\hspace{8pt}b} \theta
+i \varepsilon^{ij}
\bar{f}^{-+}_{\hspace{15pt}a} \bar{\theta} \sigma^- \theta\nonumber\\
&+&\frac{i}{4} \varepsilon^{ij} \bar{f}_{abc}
\bar{\theta} \sigma^- \sigma^{bc} \theta
-\frac{i}{12} \varepsilon^{ij}
\bar{f}_{bcd}
\bar{\theta}\sigma^- \sigma_a^{\hspace{8pt}bcd}\theta\nonumber\\
&-&i  \sqrt{-h} h^{ij}{\bar{f}}^{-+}_{\hspace{15pt}b}
\theta \sigma^- \sigma_a^{\hspace{8pt}b} \theta
+ \frac{i}{4}  \sqrt{-h} h^{ij}{\bar{f}}_{abc}
\theta \sigma^- \sigma^{bc}\theta\nonumber\\
&+&\lk.\lk.
\sqrt{-h} h^{ij}  q_a \bar{\theta} \sigma^- \theta
-  \varepsilon^{ij} q_b
{\theta} \sigma^- \sigma_a^{\hspace{8pt}b}\theta\re)+\mbox{c.c.}\re]
\label{I2}\ .
\eee
The $\sigma^a$ matrices are $16\times 16$ gamma matrices.  We denote
tangent space indices by $a,b,...$, while spacetime indices are labeled,
as in~\pref{I0}, by $m,n,...$. All tensors will be written with their
indices in the tangent space by using the vielbein
$X_{ab..}=e_a^m e_b^n X_{mn}$. The $'+'$ and $'-'$ tangent
space labels refer to the light-cone directions as in
$x^\pm\equiv (x^0\pm x^a)/2$, with $x^0$ and $x^a$ being respectively the
time and some chosen space direction defining the light-cone. 
We also have
\bb
V_i^a\equiv \del_i x^m e_m^a\ .
\ee
Furthermore, all Latin indices run over only directions
transverse to the light-cone. Finally, the covariant derivative is
given by
\bb\label{Dtheta}
\hat{D}_j \theta^\alpha\equiv \del_j \theta^\alpha 
+\frac{1}{4} \sigma^{ab\alpha}_\beta \del_j x^m
\Omega_{m,ab} \theta^\beta\ .
\ee

The various background fields appearing in~\pref{I2}, and in
subsequent equations, are:
\begin{itemize}
\item The IIB dilaton
\bb
\omega\equiv e^{\phi/2}\ .
\ee

\item The field strengths for the IIB scalars
\bb
p_m\equiv \frac{1}{2} \lk(i \hat{D}_m \chi- e^{-\phi} \hat{D}_m\phi\re)\ ;
\ee
\bb
q_m\equiv-\frac{1}{4} \hat{D}_m \chi\ ,
\ee
with $\chi$ being the IIB axion.

\item The complex field strength 
\bb
f_{abc}\equiv
\frac{1}{2}
(1+e^{-\phi}+i\chi)\FF_{abc}
+\frac{1}{2}(-1+e^{-\phi}+i\chi)\bar{\FF}_{abc}
\ ;
\ee
\bb
\FF_{abc}\equiv \frac{h^{(1)}_{abc}}{2}+i\frac{h^{(2)}_{abc}}{2}\ ,
\ee
with $h^{(1)}$ and $h^{(2)}$ being, respectively, the field
strengths associated with fundamental string and D-string charge.

\item And the five-form self-dual field strength
\bb
Z_{abcde}\equiv \frac{1}{192} g_{abcde}\ .
\ee
\end{itemize}

At quartic order in the fermions, the action involves many more terms:
\bbb
I^{(4)}&=&
\int d^2\sigma\  
\sqrt{-h} h^{ij}
(\theta \sigma^-\sigma^{a_1a_2}\theta)
(\bar{\theta} \sigma^-\sigma^{a_1a_2}\bar{\theta})
V^+_i V^+_j\times \nonumber \\
& &\lk(
-\frac{1}{16} R^{-+-+}
-72\omega^4 Z^{-+abc}Z^{-+}_{\hspace{15pt}abc}
-\frac{11}{256} \omega^4 f^{-+a} \bar{f}^{-+}_{\hspace{15pt}a}
-\frac{29}{18432} \omega^4 f^{abc} \bar{f}_{abc}
\re)\nonumber \\
&+&  \sqrt{-h} h^{ij}
(\theta \sigma^-\sigma^{b_1b_2}\theta)
(\bar{\theta} \sigma^-\sigma_{b_1}^{\hspace{8pt}b_3}\bar{\theta})
V^+_i V^+_j\times\nonumber \\
& &\lk(
-\frac{\omega^4}{8} f^{-+}_{\hspace{15pt}b_2} p_{b_3}
+\frac{5}{96} R^{-+}_{\hspace{15pt}b_2b_3}
+\frac{1}{96} R^{-\hspace{8pt}+}_{\hspace{8pt}b_2\hspace{8pt}b_3}
+\frac{23}{96} R^{-\hspace{8pt}+}_{\hspace{8pt}b_3\hspace{8pt}b_2}
\re.\nonumber \\
&+&768 \omega^4 Z^{-+ a b}_{\hspace{30pt}b_2} Z^{-+}_{\hspace{15pt}a b b_3}
\nonumber\\
&+&\frac{19}{256} \omega^4 f_{a b_2b_3} \bar{f}^{-+ a}
-\frac{55}{128}\omega^4 f^{-+}_{\hspace{15pt}b_3} 
\bar{f}^{-+}_{\hspace{15pt}b_2}
-\frac{67}{384}\omega^4 f^{-+}_{\hspace{15pt}b_2} 
\bar{f}^{-+}_{\hspace{15pt}b_3}\nonumber\\
&+&\lk.\frac{37}{3072}\omega^4 f_{a b b_3} \bar{f}^{a b}_{\hspace{15pt}b_2}
-\frac{13}{3072} \omega^4 f_{a b b_2} \bar{f}^{a b}_{\hspace{15pt}b_3}
-\frac{37}{768} \omega^4 f^{-+ a} \bar{f}_{a b_2b_3}
\re)\nonumber \\
&+&  \sqrt{-h} h^{ij}
(\theta \sigma^-\sigma^{c_1c_2}\theta)
(\bar{\theta} \sigma^-\sigma^{c_3c_4}\bar{\theta})
V^+_i V^+_j\times\nonumber \\
& &\lk(
12 i\omega^2 {\hat{D}}_{c_3} Z^{-+}_{\hspace{15pt}c_1c_2c_4}
+12 i\omega^2 \Omega_{c_3c_1c_2c_4}
+\frac{1}{32}\omega^4 f_{c_1c_2c_4} p_{c_3}
+\frac{1}{32} R_{c_1c_2c_3c_4}\re.\nonumber\\
&+&\lk.672 \omega^4 
Z^{-+ a}_{\hspace{22pt}c_1c_4} Z^{-+}_{\hspace{15pt}a c_2c_3}
-624 \omega^4 Z^{-+ a}_{\hspace{22pt}c_1c_2} 
Z^{-+}_{\hspace{15pt}a c_3c_4}\re.\nonumber\\
&+&\lk.24\omega^4 Z^{-+}_{\hspace{15pt}a b c_4} 
Z^{a b}_{\hspace{15pt}c_1c_2c_3}
-24\omega^4 Z^{-+}_{\hspace{15pt}a b c_2} 
Z^{a b}_{\hspace{15pt}c_1c_3c_4}\re.\nonumber\\
&+&\lk.\frac{31}{512}\omega^4 f_{c_2c_3c_4} \bar{f}^{-+}_{\hspace{15pt}c_1}
+\frac{61}{1536} \omega^4 f_{c_1c_2c_4} \bar{f}^{-+}_{\hspace{15pt}c_3}
\re.\nonumber\\
&-&\lk.\frac{13}{2048} \omega^4 f_{a c_3c_4} \bar{f}^{a}_{\hspace{8pt}c_1c_2}
+\frac{13}{1536} \omega^4 f_{a c_2c_4} \bar{f}^{a}_{\hspace{8pt}c_1c_3}
\re.\nonumber\\
&+&\lk.\frac{35}{2048} \omega^4 f_{a c_1c_2}\bar{f}^{a}_{\hspace{8pt}c_3c_4}
-\frac{179}{1536}\omega^4 f^{-+}_{\hspace{15pt}c_3} \bar{f}_{c_1c_2c_4}
+\frac{143}{1536}\omega^4 f^{-+}_{\hspace{15pt}c_2} \bar{f}_{c_1c_3c_4}
\re)\nonumber\\
&+&\mbox{c.c.}\label{I4}\ ,
\eee
with the same conventions as before, and
\bb\label{Rr}
R^{ab}_{\mbox{    }cd}\equiv r^{ab}_{\mbox{    }cd}
+\delta^{[a}_{[c} \Omega^{b]}_{d]}\ ,
\ee
\bb\label{rpiece}
\Omega^a_b\equiv 2({\hat{D}}^a {\hat{D}}_b \ln \omega)
+({\hat{D}}_b\ln \omega) ({\hat{D}}^a\ln \omega)
-\frac{1}{2} ({\hat{D}}^c\ln\omega) ({\hat{D}}_c\ln\omega) \delta^a_b\ ,
\ee
\bb\label{dz}
\Omega_{c_3c_1c_2c_4}\equiv 5 ({\hat{D}}_{c_3}\ln \omega) 
Z^{-+}_{\hspace{15pt}c_1c_2c_4}
-\frac{3}{2} \eta_{c_3[c_1} Z^{-+}_{\hspace{15pt}c_2c_4]b}
({\hat{D}}^b \ln \omega)\ .
\ee
$r_{abcd}$ is the Riemann tensor associated with the metric $G_{mn}$,
and the additional pieces in~\pref{rpiece} and~\pref{dz}
come from rescaling the metric from
the Einstein frame to the string frame.
More details on the notation
used can be found in the main text and in Appendix A.

The term labeled $J^{(4)}$ in equation~\pref{genform} involves interactions quartic in the fermions which are
of the form $\theta^4$ and $\theta^3 {\bar{\theta}}$ (and their complex conjugates); \ie\ terms
that carry four and
two units of U(1) charge respectively. These pieces have not been computed at the time of this writing. 
A future revision of this work will write the explicit form of $J^{(4)}$ as well to complete the action.

Equations~\pref{I2} and~\pref{I4} can be, in practice, quite simple. 
Note that some terms may drop out at the expense of others.
For example, for backgrounds which are electric or magnetic, but not
dyonic, a fraction of the terms are left: say either involving forms like
$f_{-+a}$ or $f_{abc}$, but not both simultaneously.
We will comment on some of 
the physical implications of this action in the Discussion
section. For now, let us present some of the details on how these
couplings were derived.

The outline of the paper is as follows. In Section 1, we present brief
reviews of the techniques we employ; these are the superspace formalism
for type IIB supergravity and for the IIB string sigma model, and the
normal coordinate expansion method.
In Section 2, we apply these techniques to the case of interest; first,
we present a series of arguments that simplify the discussion by
making use of the light-cone gauge and conditions imposed
on the background fields; we then outline
in some detail how to obtain the terms quadratic in the fermions; then,
more briefly, we sketch how to determine
the terms quartic in the fermions.
We end in the Discussion section with comments on the form of the
action, and preliminary remarks on how the results may be applied in
certain examples such as NCOS theories.

\section{Preliminaries}

\subsection{IIB supergravity in superspace}

The fields of
IIB supergravity are
\bb\label{fields}
\lk\{
{\hat{e}}^a_m,\  \tau=e^{-\phi}+i \chi,\  b^{(1)}_{mn} + i\  b^{(2)}_{mn},\ 
b_{mnrs};\  \psi_m,\ \lambda
\re\}\ ;
\ee
these are respectively
the vielbein, a complex scalar comprised of the
dilaton and the axion, two two-form gauge fields, a four-form real gauge field,
a complex left-handed gravitino, and a complex right-handed spinor. 
The gauge fields have the associated field strengths defined as
\bb
h^{(1)}=db^{(1)}\ ,\ \ \ 
h^{(2)}=db^{(2)}\ ,\ \ \ 
g=db\ .
\ee

An elaborate superspace formalism can be developed for this theory.
It involves the standard supergravity superfields~\cite{HOWEWEST}
\bb\label{superfields1}
\lk(E^A_M,\Omega_{MA}^{B}\re)\rightarrow
\lk(T^A_{BC},R_{ABC}^{D}\re)\ .
\ee
In addition, one needs five other tensor superfields
\bb\label{superfields2}
\lk\{P_A,Q_A,{\hat{\FF}}_{ABC},G_{ABCDE},\Lambda_A\re\}
\ee
Throughout, we accord
to the standard convention of denoting tangent space
superspace indices by capital
letters from the beginning of the alphabet. 
In this setting of $\NN=2$ chiral supersymmetry, 
an index such as $A$ represents a tangent space 
vector index $a$, and two spinor
indices $\alpha$ and $\bar{\alpha}$.
Hence, superspace is
parameterized by coordinates
\bb
z^A\in\lk\{x^a,\theta^\alpha,\theta^{\bar{\alpha}}\re\}\ .
\ee
Here, $\theta^\alpha$ and $\theta^{\bar{\alpha}}\equiv {\bar{\theta}}^\alpha$
have same chirality and are related to each other by complex conjugation. 
In this manner, 
unbarred and barred Greek letters from the beginning of the alphabet
will be used to denote spinor indices. 
More details about the conventions we adopt can be found in Appendix A.

The two superfields $P_A$ and $Q_A$ are the field strengths of a matrix
of scalar superfields
\bb
\VV=\lk(
\begin{array}{cc}
u & v \\
\overline{v} & \overline{u}
\end{array}
\re)
\ ,
\ee
with
\bb
u\overline{u}-v\overline{v}=1\ .
\ee
This matrix describes the group $SU(1,1)\sim SL(2,R)$, which
later gets identified with the S-duality group of the IIB theory.
The scalars parameterize the coset space $SU(1,1)/U(1)$, with the additional
$U(1)$ being a space-time dependent symmetry with an associated gauge field.
We then define
\bb
\VV^{-1} d\VV\equiv \lk(
\begin{array}{cc}
2 i Q & P \\
\overline{P} & -2 i Q
\end{array}
\re)\ ,
\ee
with
\bb
Q=\overline{Q}
\ee
being the $U(1)$ gauge field mentioned above. All fields in the theory
carry accordingly various charge assignments under this $U(1)$. This is
a powerful symmetry that can be used to severely restrict the superspace
formalism.
We also introduce the superfield strength $\hat{F}$
\bb
\lk( \bar{{\hat{\FF}}}, \hat{\FF}\re)=
\lk(\bar{{\hat{F}}},\hat{F}\re)\VV^{-1}\ ,
\ee
which transforms under the $SU(1,1)$ as a singlet.

All these fields are associated with a myriad of Bianchi identities.
As is typical in supergravity theories, there is an immense amount of
superfluous symmetries in the superspace formalism. Some of these
can be fixed conventionally; and using the Bianchi identities, relations
can be derived relating the various other components. 
We will be very brief in reviewing this formalism, as 
our focus will be the string sigma model. Instead of reproducing
the full set of equations that determine the IIB theory, we present 
only those statements that are of direct relevance to the
worldsheet theory.
Throughout this work,
we accord closely to the conventions and notation
of \cite{HOWEWEST}; the reader may refer
at any point to that work to complement his/her understanding.

From the point of view of the IIB string sigma model, the following
combination of the scalars turns out to play an important role
\bb
\omega=u-\overline{v}\ .
\ee
Requiring $\kappa$ symmetry on the worldsheet leads to the condition
\bb\label{gauge}
\omega=\bar{\omega}\ .
\ee
This is a choice that is unconventional from the point of view
of the supergravity formalism, but it is natural from the perspective
of the string sigma model.

We parameterize the scalar superfields as~\cite{RADAK,BELLUCI}
\bb
u=\frac{1+\bar{W}}{\sqrt{2 (W+\bar{W})}} e^{-2 i \theta}\ ,
\ee
\bb
v=-\frac{1-W}{\sqrt{2 (W+\bar{W})}} e^{2 i \theta}\ ,
\ee
with the three variables $W$, $\bar{W}$ and $\theta$ parameterizing the
$SU(1,1)$. The gauge choice~\pref{gauge} then corresponds to
\bb
\theta=0\ ,
\ee
This leads to
\bb
\omega=\sqrt{\frac{2}{W+\bar{W}}}\ .
\ee
And
\bb
Q_A=\frac{\bar{P}_A-P_A}{4 i}\ .
\ee
Finally, the field strengths are given in terms of $W$ by
\bb
P=\frac{dW}{W+\bar{W}}\ ,\ \ \ 
Q=\frac{i}{4} \frac{d(W-\bar{W})}{W+\bar{W}}\ .
\ee

To make contact with the IIB theory's field content~\pref{fields},
we need to specify the map between the superfields~\pref{superfields1} and
~\pref{superfields2} and the
physical fields. Each superfield involves an expansion in the fermionic
superspace coordinates $\theta$. At zeroth order in this expansion, we 
have
\bb
W|_0=\tau=e^{-\phi}+i \chi\ ,
\ee
Similarly, the zeroth components of the $\Lambda$ superfield is
\bb
\Lambda_\alpha|_0=\lambda_\alpha\ .
\ee
In the Wess-Zumino gauge, the supervielbein's zeroth component is
\bb
E_{m}^\alpha|_0=\psi_m^\alpha\ .
\ee
At this point, we can simplify the discussion significantly by 
choosing to set all background fermionic fields to zero
\bb
\lambda_\alpha\rightarrow 0\ ,\ \ \ 
\psi_m^\alpha\rightarrow 0\ .
\ee
This identifies the class of backgrounds which is of most interest
to us and  that arises most frequently in the literature.
Given this, the zeroth components of the other fields are
\bb
{\hat{\FF}}_{abc}|_0=\FF_{abc}\equiv
\frac{h^{(1)}_{abc}}{2}
+i \frac{h^{(2)}_{abc}}{2}\ ,
\ee
\bb
G_{abcde}|_0=g_{abcde}\ .
\ee
We also define
${\hat{F}}_{abc}|_0\equiv F_{abc}$.
And, for completeness, we write the full form of the supervielbein
\bb
E^A_M|_0=\lk(
\begin{array}{ccc}
{\hat{e}}^a_m & 0 & 0\\
0 & \delta^\alpha_\mu & 0 \\
0 & 0 & -\delta^{\bar{\alpha}}_{\bar{\mu}}
\end{array}
\re)\ ;
\ee
with the zeroth components of the connection
\bb
\Omega_{cA}^B|_0=
\omega_{c,A}^B
+\mbox{U(1) connection}\ ;
\ee
\bb
\Omega_{\alpha,A}^B|_0=
\Omega_{\bar{\alpha},A}^B|_0=0\ ;
\ee
and the other combinations of indices being zero.

For reasons that will become apparent later, the fields appearing
in~\pref{I2} and~\pref{I4} are further
rescaled with respect to the ones presented here,
as in
\bb\label{redef}
F_{abc}\equiv\omega f_{abc}\ ,\ \ \ 
Q_a|_0\equiv \omega^2 q_a\ ,\ \ \ \ 
P_a|_0\equiv \omega^2 p_a\ .
\ee

In addition, we will need the zeroth components of the
Riemann and torsion superfields, as well as various spinorial components
of all the superfields. To make things even worse, 
various first and second order
spinorial derivatives of the superfields will also be needed; 
\ie\ some of the higher order
terms in the superfield expansions appear in the sigma model. 
These can be systematically, albeit sometimes tediously,  obtained
by juggling the superspace 
Bianchi identities. The needed set turns out
not to be exhaustive in terms of determining the IIB supergravity.
We will present the relevant  
pieces as we need them, instead of cataloging an incomplete set of 
lengthy equations out of context.

Finally, to relate these fields to the ones that arise
in the modern literature,
we write down the equations of motion of the zero component fields
as they appear in the relations above:
\bb
\lk(\tau+\bar{\tau}\re) \nabla^2 \tau-2\lk(\nabla\tau\re)^2
+\frac{1}{12} \lk(\tau+\bar{\tau}\re)\lk(h^{(2)}-i \tau h^{(1)}\re)^2=0\ ;
\ee
\bb
\nabla^p\lk(
\frac{2\tau\bar{\tau} h^{(1)}_{mnp}
+i (\bar{\tau}-\tau) h^{(2)}_{mnp}}{\tau+\bar{\tau}}
\re)+\frac{1}{6}g^{abcde} h^{(2)}_{cde}=0\ ;
\ee
\bb
\nabla^p\lk(
\frac{2 h^{(2)}_{mnp}
+i (\bar{\tau}-\tau) h^{(1)}_{mnp}}{\tau+\bar{\tau}}
\re)-\frac{1}{6} g^{abcde} h^{(1)}_{cde}=0\ ;
\ee
which conform, for example,
to those in~\cite{SL2Z} with $i\bar{\tau}\rightarrow\lambda$, 
$h^{(1)}\rightarrow H^{(1)}$,
$h^{(2)}\rightarrow H^{(2)}$.

\subsection{The IIB string worldsheet in superspace}

The action of the IIB string in a background represented
by the superfields listed above was written in~\cite{GHMNT}
\bb\label{susyaction}
I=\int d^2\sigma\ \lk\{
\frac{1}{2} \sqrt{-h} h^{ij} \Phi V_i^a V_j^b \eta_{ab}
+\frac{1}{2} \varepsilon^{ij} V_i^B V_j^A \BB_{AB}
\re\}\ ,
\ee
with
\bb
V_i^A\equiv \del_i z^M E_M^A=
\lk\{V_i^a,V_i^\alpha,V_i^{\bar{\alpha}}
\re\}\ ,
\ee
and
\bb
d\BB=\hat{\FF}+\bar{{\hat{\FF}}}\ ,
\ee
\bb
\Phi=\omega=\bar{\Phi}\ .
\ee
The last statement is needed to assure that the action
is $\kappa$ symmetric.
The task is to expand this action in component form. This is generally
a messy matter, which, however, can be achieved using the
algorithm of normal coordinate expansion.

\subsection{The method of normal coordinate expansion in superspace}

Normal coordinate expansion, as applied to bosonic sigma models,
was first developed in~\cite{MUKHINLSM}. In these scenarios, the method helped
to unravel some of
the dynamics of highly non-linear theories approximately, 
as expanded near a chosen point on the target manifold.
In the superspace incarnation, the technique is most
powerful when used to expand 
an action only in a submanifold of the target superspace. In particular,
expanding in the fermionic variables only, with the space coordinate 
left arbitrarily, the expansion truncates by virtue of the Grassmanian
nature of the fermionic coordinates; leading to an exact expression
for the action in component form. This can also be applied of course to
the action or equations of motion
for the background superfields as well, and the technique 
has been demonstrated in this context in many examples. As for the 
IIB sigma model, the expansion has been applied in~\cite{GMPRV}, to 
expand however
the action in all of superspace, leading to a linearized approximate
form that can be used to study quantum effects. Our interest is to
get to an exact expression for~\pref{susyaction} in component form, by fixing
the $\kappa$ symmetry and leaving the space coordinates arbitrary. 
This approach was applied to the Heterotic string in~\cite{ATICKDHAR}. 
There, the absence of RR fields made the discussion considerably 
simpler. Our approach will probe in this respect a new class of couplings
by the use of this method. However, many simplifications and techniques
we will use are direct generalizations of the corresponding 
methods applied in~\cite{ATICKDHAR}. 
First, we briefly review the normal coordinate
expansion method in superspace. The reader is referred 
to~\cite{GRISARU,ATICKDHAR}
for more information.

The superspace coordinates are written as
\bb
Z^M=Z_0^M+y^M\ .
\ee
We choose
\bb
Z_0^M=\lk(x^m,0\re)\ ,\ \ \ 
y^M=\lk(0,y^\mu\re)\ ,
\ee
hence expanding only in the fermionic submanifold.
The action is then given by 
\bb\label{edelta}
I[Z]=e^{\Delta} I[Z_0]\ ,
\ee
with the operator $\Delta$ defined by
\bb
\Delta\equiv\int d^2 \sigma\ y^A(\sigma) D_A(\sigma)\ ,
\ee
and $D_A$ being the supercovariant derivative.
Here, we use the supervielbein to translate between tangent space
and superspacetime indices
\bb
D_A\equiv E^N_A(Z_0) D_N\ ,\ \ \ 
y^N\equiv y^A E_A^N\ .
\ee
For our choice of expansion variables, we then have
\bb
y^a=0\ ,\ \ \ 
y^\alpha=y^\mu\delta_\mu^\alpha\equiv \theta^\alpha\ ,\ \ \ 
y^{\bar{\alpha}}=y^{\bar{\mu}}\delta_{\bar{\mu}}^{\bar{\alpha}}\equiv 
\theta^{\bar{\alpha}}\ .
\ee

The power of this technique is that it renders the process of expansion
{\em algorithmic}. A set of rules
can be taught say to any well-trained mammal; 
in principle, human intervention
(for that matter the same mammal may be used again) is needed only
at the final stage when
Bianchi identities may be used to determine some of
the expansion terms. The rules are as follows:

\begin{itemize}
\item Due to the definition of the normal coordinates, we have
\bb\label{rule1}
\Delta y^A=0\ .
\ee

\item Using super-Lie derivatives, it is straightforward to derive
\bb
\Delta V_i^A=D_i y^A+V_i^C y^B T_{BC}^A\ .
\ee

\item And the following identity is needed beyond second order
\bb
\Delta\lk(D_iy^A\re)=y^B V_i^D y^C R_{CDB}^{A}\ .
\ee

\item Finally, when we apply $\Delta$ to an arbitrary tensor with
tangent space indices, we get simply
\bb\label{rule4}
\Delta X_{BC..}^{DE..}=y^A D_A X_{BC..}^{DE..}\ .
\ee
\end{itemize}
In the next section, we outline the process of applying these rules 
to~\pref{susyaction}.

\section{Unraveling the action}

There are three sets of difficulties that arise when attempting to
apply the normal coordinate expansion to~\pref{susyaction}. First, a priori, 
we need to expand to order $2^{5}$ in $\theta$ before the expansion
truncates. This problem 
is remedied simply by fixing the $\kappa$ symmetry
with the light-cone gauge, truncating the action to quartic
order in $\theta$, as we will show below.
The second problem is that the expansion terms will need first and second
order fermionic derivatives of the superfields. This requires us to play around
with some of the Bianchi identities to extract the additional information.
The process is somewhat tedious, but straightforward.
The third problem is computational. Despite the simplifications induced
by the light-cone gauge choice, and the algorithmic nature of the process,
it turns out that the task is virtually impossible to perform by a human,
while still maintaining some level of
confidence in the result. On average $10^4$
terms arise at various stages of the computation. The use of the computer
for these analytical manipulations simplifies matters further. However,
we find that, even with this help, the complexity is large enough that computing
time is of order of many weeks, unless the task is approached with a set
of somewhat smarter computational steps and unless one uses
the simplifications that arise from the
conditions imposed on the background fields and listed in the Introduction.
We do not present all the messy
details of these nuances, concentrating instead on the general protocol.

At zeroth order, the action is simply
\bb
I^{(0)}=I|_0=\int d^2 \sigma \lk\{
\frac{1}{2} \sqrt{-h} h^{ij}\omega V_i^a V_{ja}
+\frac{1}{2} \varepsilon^{ij}  V_i^b V_j^a b_{ab}^{(1)}
\re\}\ .
\ee
Note that this is written with respect to the Einstein frame
metric, as determined by the IIB supergravity formalism presented
in the previous section. We will rescale it to the string frame at the end.

At first order in $\Delta$, the action becomes
\bbb
I^{(1)}&=&\Delta I=
\int\ d^2\sigma \lk\{
\frac{1}{2} \sqrt{-h} h^{ij} (\Delta\Phi) V_i^a V_j^b \eta_{ab}
+\sqrt{-h} h^{ij} \Phi (\Delta V_i^a) V_j^b \eta_{ab}\re.\nonumber \\
&+&\lk.\frac{1}{2} \varepsilon^{ij} V_i^B V_j^A y^C \HH_{CBA}
\re\}\label{deltaI}\ ,
\eee
with
\bb
\HH\equiv d\BB\ .
\ee
This result is not evaluated at zeroth order in $\theta$ yet, as
further powers of $\Delta$ will hit it. The rest is mostly 
mechanical.

\subsection{Fixing the $\kappa$ symmetry}

It simplifies matters if we analyze the form of the action we
expect from this expansion, once the $\kappa$ symmetry is fixed.
This will help us avoid manipulating many of the terms that will
turn out to be zero in the light-cone gauge. To fix the $\kappa$
symmetry, we define
\bb\label{sigmapm}
\sigma^{\pm}\equiv \frac{1}{2} \lk(\sigma^0\pm \sigma^a\re)\ ,
\ee
where $a$ is some chosen direction in space.
For conventions on spinors, the reader is referred to Appendix A
and~\cite{HOWEWEST}. We choose the spacetime fermions to satisfy the condition
\footnote{
Alternatively, we can choose~\cite{PESANDO}
\bb
\sigma^{\pm}\equiv \frac{1}{2} \lk(\sigma^a\pm i \sigma^b\re)\ ,
\ee
with $a$ and $b$ being two arbitrary space directions. We can then impose
\bb
\sigma^+\theta=\sigma^-\bar{\theta}=0\ .
\ee
It can be seen that this choice leads to a more complicated
expansion for the action. It may still be necessary to consider 
such choices for other classes of background fields than those
we focus on in this work.
}
\bb\label{LC}
\sigma^+\theta=\sigma^+\bar{\theta}=0\ .
\ee

Consider first all even powers of $\theta$. These will necessarily 
come within one of the following bilinear combinations
\bb\label{bi1}
A^{ab}\equiv \theta \sigma^-\sigma^{ab}\theta\ ,\ \ \ 
\bar{A}^{ab}=\bar{\theta}\sigma^-\sigma^{ab}\bar{\theta}\ ;
\ee
\bb\label{bi2}
B\equiv\bar{\theta} \sigma^- \theta\ ,\ \ \ 
B^{ab}\equiv\bar{\theta}\sigma^-\sigma^{ab}\theta\ ,\ \ \ 
B^{abcd}\equiv \bar{\theta}\sigma^-\sigma^{abcd}\theta\ .
\ee
In these expressions, condition~\pref{LC} has been used, and 
the Latin indices are necessarily transverse to the light cone
directions. We will not keep track of this fact with any special
notation as it should be obvious from the context where it arises.
Given the symmetry properties of the gamma matrices (see Appendix A),
we also have
\bb
\bar{B}=B\ ,\ \ \ 
\bar{B}^{ab}=-B^{ab}\ ,\ \ \ 
\bar{B}^{abcd}=B^{abcd}\ .
\ee

%We then can contract pairs of spinors and
%generate relations such as
%\bb
%A^{ab} A^{cd}=-\frac{1}{16} \frac{1}{32} A^{ef} A^{gh} \mbox{Tr}
%[\sigma^{ab} \sigma_{ef} \sigma^{cd} \sigma_{gh}]\ .
%\ee
%Another example is
%\bb
%A^{ab} B=-\frac{1}{16} A^{cd} \lk(
%\frac{1}{32} B^{ef} \mbox{Tr} [\sigma_{cd} \sigma^{ab} \sigma_{ef}]
%-\frac{1}{16} B \mbox{Tr} [\sigma_{cd} \sigma^{ab}]
%-\frac{1}{16} \frac{1}{4!} B^{efgh} \mbox{Tr} 
%[\sigma_{cd} \sigma^{ab} \sigma_{efgh}]\re)\ .
%\ee
%Using such relations, it is straightforward to show that
%\bb
%A^{ab} A^{cd}=0\ ,\ \ \ 
%\bar{A}^{ab} \bar{A}^{cd}=0\ .
%\ee
%And
%\bb
%B A^{ab}=0\ ,\ \ \ 
%B^{ab} A^{cd}=0\ ,\ \ \ 
%B^{abcd} A^{ef}=0\ ;
%\ee
%\bb
%B \bar{A}^{ab}=0\ ,\ \ \ 
%B^{ab} \bar{A}^{cd}=0\ ,\ \ \ 
%B^{abcd} \bar{A}^{ef}=0\ .
%\ee
%While we have
%\bb
%A^{ab} \bar{A}^{cd}\neq 0\ .
%\ee
%These immediately imply that all powers of $\theta$ higher and including
%five are zero. This does not yet truncate the action to quartic order.
%One more piece of argument is needed.

\subsection{The expected form of the action}

First, we note that, given that all background fermions ($\lambda$ and
$\psi_m$) are zero, only even powers of $\theta$ can appear in the
expansion. 
Next, we assume that all background fields have only non-zero
components that are either transverse to the light-cone directions, 
or that the light-cone indices in them
come in pairs; and that all the fields depend
only on the transverse coordinates.  For example, denoting the light-cone
directions by $'+'$ and $'-'$, and all transverse coordinates schematically
by $r$,
all fields can only depend on $r$; and a tensor $X_{abc..}$
can be non-zero only if either all $a,b,c,..$ are transverse; or if
$'+'$ and $'-'$ come as in $X_{-+bc..}$ with $b,c..$ transverse or other
light-cone pairs. 
These conditions are satisfied
by all backgrounds of particular interest to us.
And it leads to a dramatic simplification of the expansion.
In particular, given that a $'-'$ index is to appear in all even powers of
fermion bilinears, as in~\pref{bi1} and~\pref{bi2}, we must pair each bilinear
with a $V_i^a$ to absorb the light-cone index $'-'$. 

Let $\Theta$ represent either $\theta$ or $\bar{\theta}$. For example,
schematically
$\Theta^2\sim \theta^2, \bar{\theta} \theta, \bar{\theta}^2$.
The action consists then of 
terms of form 
$\Theta^{2n} V_i^a V_j^b$, 
$(D\Theta) \Theta^{2n-1} V_i^a$ and
$(D\Theta) (D\Theta) \Theta^{2n}$.
From the expansion algorithm outlined above, with the use of equations
~\pref{rule1}-\pref{rule4}, it is easy to see that
\bbb
\mbox{number of V's}+\mbox{number of }D\Theta\mbox{'s}=2\nonumber
\eee
in each term.
Let's then look at each class of terms separately:
\begin{itemize}
\item For terms of the form
$\Theta^{2n} V^a V^b$, the 
only non-zero combinations are $\Theta^2 V_i^+ V_j^a$ 
and $\Theta^4 V_i^+ V_j^+$.
This means in particular
that the Wess-Zumino term involving $\HH$ in~\pref{deltaI} does not 
contribute at quartic order since we must contract $V_i^+ V_j^+$ by $\sqrt{-h} h^{ij}$. 

\item Terms of the form
$(D\Theta) \Theta^{2n-1} V_i^a$ are zero unless $n=1$, because, otherwise,
there is shortage of $V$s
to absorb all light-cone indices.

\item Terms of the form
$(D\Theta) (D\Theta) \Theta^{2n}$ are zero for all $n$ for the same
reason as above.
\end{itemize}

Hence, the action must have the form
\bb
I\sim \Theta D\Theta+\Theta^2+\Theta^4 V^+ V^+\ ,
\ee
with the quartic piece receiving contributions only from the first
two terms of~\pref{deltaI}.
And we focus on expanding only the relevant parts.

Let us also note that for a typical class of D-brane backgrounds, the natural
choice that leads to the simplifications we outlined corresponds to
aligning the light-cone direction parallel to the worldvolume of the D-brane.
The transversality condition on the fields is then satisfied. 

Hence, the action truncates at quartic order in the fermions. In this work, we compute part of
the quartic interactions - ones of the form $\theta^2 {\bar{\theta}}^2$ carrying zero U(1) charge - and 
leave the remaining pieces for a future update of the manuscript. In the original version of this preprint, an
erroneous argument involving gamma matrix algebra lead us to believe that no additional
quartic terms would be present other than the ones that we already compute. 
This is not the case, and we can indeed expect that the action presented as
of this writing is incomplete and will involve additional pieces of the form
$\theta^4$ and $\theta^3 \bar{\theta}$ (and their complex conjugates). Yet, as our argument here 
still shows, the
action truncates to quartic order irrespective of this issue.

\subsection{The quadratic terms}

As we expand~\pref{edelta}, the quadratic terms in $\theta$ are very simple 
to handle, and can be done by hand. On finds that zeroth components
of $D\omega$ and $D^2 \omega$ are needed. For these, we note the relation
\bb
d\omega=-\frac{\omega}{2} \lk(P+\bar{P}\re)\ .
\ee
Using the results of~\cite{HOWEWEST}, we get
\bb
D_\alpha \omega|_0=D_{\bar{\alpha}}\omega |_0=0\ .
\ee
\bb
D_{\alpha} D_{\beta} \omega|_0=-\omega 
\frac{i}{24} \sigma^{abc}_{\alpha\beta} F_{abc}\ ,\ \ \ 
D_{\bar{\alpha}} D_{\bar{\beta}} \omega|_0=-\omega 
\frac{i}{24} \sigma^{abc}_{\alpha\beta} \bar{F}_{abc}\ ,
\ee
\bb
D_{\bar{\alpha}} D_{\beta} \omega|_0=-\omega 
\frac{i}{2} \sigma^a_{\alpha\beta} P_a|_0\ ,\ \ \ 
D_{{\alpha}} D_{\bar{\beta}} \omega|_0=-\omega 
\frac{i}{2} \sigma^a_{\alpha\beta} \bar{P}_a|_0\ .
\ee
Note that the supercovariant derivative $D_A$ is associated
with the standard supergravity superconnection plus the
$U(1)$ piece, as discussed in~\cite{HOWEWEST}.
In these and subsequent equations, a Latin index runs over all directions,
the transverse and the light-cone ones.
In the Wess-Zumino term, we need
$D_\alpha \HH_{\beta ab}|_0$ and
$D_{\bar{\alpha}} \HH_{\beta ab}|_0$.
These are found
\bb
D_\alpha \HH_{\beta ab}|_0=i\frac{\omega}{2}
\sigma_{ab\beta}^{\hspace{20pt}\gamma}
\sigma^c_{\alpha\gamma} \bar{P}_c|_0
\ee
\bb
D_{\bar{\alpha}} \HH_{\beta ab}|_0=i\frac{\omega}{24}
\sigma_{ab\beta}^{\hspace{20pt}\gamma} 
\sigma^{cde}_{\bar{\alpha}\gamma} \bar{F}_{cde}\ .
\ee

The result of applying these relations can be grouped into
several parts.
A Weyl term
\bb
I_{Weyl}=\frac{i}{96} \sqrt{-h} h^{ij} \omega F_{abc}
\theta\sigma^{abc}\theta V_{dj} V^{d}_{i}
-\frac{i}{8} \sqrt{-h} h^{ij} \omega 
P_a \bar{\theta}\sigma^a\theta V_{bi} V^b_j\rightarrow 0
\ee
is zero by the transversality condition on the background fields.
The kinetic term takes the standard form
\bb\label{Ikin}
I_{Kin}=-\frac{i}{2} \omega V_{ai}
\lk(
\varepsilon^{ij} \theta \sigma^a D_j\theta
+\sqrt{-h} h^{ij} \bar{\theta}\sigma^a D_j\theta
\re)\ ,
\ee
with
\bb
D_m \theta^\alpha\equiv \del_m \theta^\alpha +\frac{1}{4}
\sigma_\beta^{ab\alpha}\omega_{m,ab} \theta^\beta 
+i Q_m \theta^\alpha\ .
\ee
And the quadratic terms look like
\bbb
I_{quad}&=&2\omega Z_{abcde} \Sigma_1^{cde,ij} V_{i}^a V_{j}^b
-\frac{\omega}{2} Z_{bcdef} \Sigma_1^{acdef,ij} V_{ai} V^{b}_j\nonumber\\
&+&\frac{i}{8}\omega {\bar{F}}_{bcd} \Sigma_2^{acd,ij} V_{ai} V^b_j
-\frac{i}{8}\omega {\bar{F}}_{abc} \Sigma_2^{c,ij} V^a_i V^b_j
-\frac{i}{48} \omega {\bar{F}}_{cde} \Sigma_2^{abcde,ij} V_{ai} V_{bj}
\nonumber\\
&+&I_P+\mbox{c.c}
\eee
with
\bb\label{Ip}
I_P\equiv 
\frac{i}{8}\omega \bar{P}_c|_0 \Sigma_1^{abc,ij} V_{ai} V_{bj}\ .
\ee
And we have defined
\bb
\Sigma_1^{(r),ij}\equiv
\varepsilon^{ij} \theta \sigma^{(r)} \theta
+\sqrt{-h} h^{ij} \bar{\theta}\sigma^{(r)} \theta\ ;
\ee
\bb
\Sigma_2^{(r),ij}\equiv \varepsilon^{ij} \bar{\theta}
\sigma^{(r)} \theta +\sqrt{-h} h^{ij} {\theta} \sigma^{(r)} {\theta}\ .
\ee
We separated the piece in~\pref{Ip} since it involves a coupling
to the gradient of the dilaton. This term can be absorbed
into the connection by rescaling the metric from the Einstein frame to
the string frame
\bb
{{e}}^a_m=\sqrt{\omega} {\hat{e}}^a_m\Rightarrow
{G}_{mn}=\omega g_{mn}
\ .
\ee
The covariant derivative then becomes
\bb
D_j\theta^\alpha=
\hat{D}_j\theta^\alpha
+\frac{1}{4} \sigma^{ab\alpha}_\beta \theta^\beta V_{aj}
\del_b \lk(\ln \omega \re) 
+i V_j^a Q_a \theta^\alpha\ ,
\ee
with $\hat{D}$ defined in~\pref{Dtheta}. In that equation,
$\Omega_{m,ab}$ is the connection associated with the string frame
metric $G_{mn}$.
Massaging these equations back into~\pref{Ikin}, and making use of
the fact that the complex conjugate is also to added to everything
to make the action real, we get
\bbb
I_{kin}+I_P&=&
-\frac{i}{2} \omega V_{ai} \lk(
\varepsilon^{ij} \theta \sigma^a \hat{D}_j \theta
+\sqrt{-h} h^{ij} \bar{\theta}\sigma^a \hat{D}_j \theta
\re)\nonumber \\
&-&\frac{\omega}{4} Q_c|_0 \Sigma_1^{abc,ij} V_{ai} V_{bj}
+\frac{\omega}{2} V_{ai} V^b_j Q_b|_0 \sqrt{-h} h^{ij}
\bar{\theta}\sigma^a\theta\ .
\eee
This is not yet the final form. Each tensor involves powers of the
vielbein, as in
\bb
X_{abc..}={\hat{e}}^m_a {\hat{e}}^n_b {\hat{e}}^p_c X_{mnp...}\ ,
\ee
and hence there are additional powers of $\omega$ from these as the vielbein
is rescaled. In particular, we have
$Z\rightarrow \omega^{5/2} Z$, $F\rightarrow \omega^{3/2} F$,
$P\rightarrow \omega^{1/2} P$, and $Q\rightarrow \omega^{1/2} Q$.
Also, each $V_i^b$ absorbs a $\sqrt{\omega}$, \ie\
$V^a\rightarrow \omega^{-1/2} V^a$.
Finally, we normalize the kinetic term by rescaling $\theta$ as
$\omega^{1/4} \theta\rightarrow \theta$. This does not
introduce any derivatives of the dilaton because of the symmetry 
properties of the gamma matrices and the form of the action. The final
step involves rewriting some of the fields as in~\pref{redef}, 
so as to express them 
in tune with more conventional choices for 
the IIB fields. Equations~\pref{I0}-\pref{I4} have these changes
applied to them; and the indices have been expanded so that
all Latin labels on tensors refer to
transverse directions only. Note that some of our choices of field 
redefinitions
are different from the ones used in~\cite{ATICKDHAR} in the case of the
Heterotic string.

\subsection{The quartic terms}

At quartic order in $\theta$, the action is much more difficult 
to find. Indeed, the use of computation by machine becomes
particularly helpful. We do not present all the details, but only
the relations that are needed to check the results.

First derivatives of some of the Riemann tensor components arise; 
particularly, 
$D_{\bar{\alpha}} {\hat{R}}_{\beta a\gamma_1}^{\gamma_2}$ and
$D_{\alpha} {\hat{R}}_{\bar{\beta} a\gamma_1}^{\gamma_2}$.
Using the results of~\cite{HOWEWEST}, it is straightforward to find
\bb
D_{\alpha} {\hat{R}}_{\bar{\beta} a\gamma_1}^{\gamma_2}|_0=
\frac{i}{8} \sigma^{cd\gamma_2}_{\gamma_1} 
\lk(
\sigma_{a\bar{\beta}\delta}D_{\alpha}T_{cd}^\delta+
\sigma_{c\bar{\beta}\delta}D_{\alpha}T_{ad}^\delta+
\sigma_{d\bar{\beta}\delta}D_{\alpha}T_{ca}^\delta
\re)|_0
+\frac{i}{2} \delta_{\gamma_1}^{\gamma_2} P_a \sigma^b_{\alpha\bar{\beta}}
\bar{P}_b|_0\ ;
\ee
\bb
D_{\bar{\alpha}} {\hat{R}}_{{\beta} a\gamma_1}^{\gamma_2}|_0=
-\frac{i}{8} \sigma^{cd\gamma_2}_{\gamma_1} 
\lk(
\sigma_{a{\beta}\bar{\delta}}D_{\bar{\alpha}}T_{cd}^{\bar{\delta}}+
\sigma_{c{\beta}\bar{\delta}}D_{\bar{\alpha}}T_{ad}^{\bar{\delta}}+
\sigma_{d{\beta}\bar{\delta}}D_{\bar{\alpha}}T_{ca}^{\bar{\delta}}
\re)|_0
-\frac{i}{2} \delta_{\gamma_1}^{\gamma_2} {\bar{P}}_a \sigma^b_{\bar{\alpha}{\beta}}
{P}_b|_0\ .
\ee
We note the distinction between $R$ and $\hat{R}$; the latter includes
the curvature from the $U(1)$ gauge field, as defined in~\cite{HOWEWEST}.
To avert confusion, we also note that the covariant derivative $D_A$ is
with respect to $\hat{R}$; whereas the one appearing in the 
introduction as $\hat{D}$ does not involve the $U(1)$ connection and it
is the derivative with respect to the {\em string frame} metric, as
mentioned earlier. This aspect of our notation then differs slightly 
from that of~\cite{HOWEWEST}.

We need a series of first spinorial derivatives of the torsion.
For these, we need to use the Bianchi identity
\bb
\sum_{(ABC)} D_A T_{BC}^D
+T_{AB}^E T_{EC}^D
-{\hat{R}}_{ABC}^D=0\ ,
\ee
where the sum is over graded cyclic permutations.
We then find 
\bb
D_\alpha T_{cd}^\delta|_0=
R_{cd\alpha}^\delta
-D_dT_{\alpha c}^\delta-D_cT_{d\alpha}^\delta
+2 T_{\alpha [d}^{\bar{\beta}} T_{c]\bar{\beta}}^\delta
-2 T_{\alpha [d}^{{\beta}} T_{c]{\beta}}^\delta
+\delta_{\alpha}^{\delta} {\bar{P}}_{[c} P_{d]}\ ,
\ee
and
\bb
D_{\bar{\alpha}} T_{bc}^{\bar{\delta}}|_0=
-D_bT_{c\bar{\alpha}}^{\bar{\delta}}
-D_cT_{\bar{\alpha}b}^{\bar{\delta}}
+R_{bc\bar{\alpha}}^{\bar{\delta}}
+2 T_{\bar{\alpha}[c}^{\bar{\gamma}}T_{b]\bar{\gamma}}^{\bar{\delta}}
-2 T_{\bar{\alpha}[c}^{{\gamma}}T_{b]{\gamma}}^{\bar{\delta}}
+\delta_{\bar{\alpha}}^{\bar{\delta}} {\bar{P}}_{[b} P_{c]}\ .
\ee
We also have
\bb
D_{\alpha} T_{\bar{\beta} \bar{\gamma}}^{\delta}|_0=
-\frac{i}{24} \sigma^d_{\bar{\beta}\bar{\gamma}}
\sigma_{d}^{\delta\beta}
\sigma^{abc}_{\alpha\beta} F_{abc}
+\frac{i}{24} \delta^{\delta}_{\bar{\beta}}
\sigma^{abc}_{\alpha\bar{\gamma}} F_{abc}
+\frac{i}{24} \delta^{\delta}_{\bar{\gamma}}
\sigma^{abc}_{\alpha\bar{\beta}} F_{abc}\ .
\ee
In all these and subsequent equations, the right hand sides are to be
evaluated as zeroth order in $\theta$.

As if first derivatives are not enough of a mess, two derivatives
of the torsion are also needed. For example,
$D_\alpha D_\beta T_{\bar{\gamma},a}^{\delta}$ arises and is found
\bbb
& &D_\alpha D_\beta T_{\bar{\gamma} a}^{\delta}|_0=
-\frac{3}{16}\sigma_{\bar{\gamma}}^{de\delta}\lk(
-\frac{1}{32} 
\KK_{ade\beta}^{\hspace{20pt}\gamma} D_{\alpha}D_\gamma\omega
+3 P_{[a}\sigma_{de]\beta}^{\hspace{18pt}\gamma} D_\alpha D_\gamma\omega
+3 i \sigma_{[a\beta\gamma}D_\alpha T_{de]}^\gamma
\re)\nonumber\\
&-&\frac{1}{48} \sigma_{a\bar{\gamma}}^{\hspace{8pt}cde\delta}\lk(
-\frac{1}{32}
\KK_{cde\beta}^{\hspace{20pt}\gamma} D_{\alpha}D_\gamma\omega
+3 P_{[c}\sigma_{de]\beta}^{\hspace{18pt}\gamma} D_\alpha D_\gamma\omega
+3 i \sigma_{[c\beta\gamma}D_\alpha T_{de]}^\gamma
\re)\ ,
\eee
where we define the matrix
\bb
\KK_{cde}\equiv \sigma_{cdefgh} \bar{F}^{fgh}
+3 \bar{F}_{[c}^{\hspace{7pt}fg}\sigma_{de]fg}
+52 \bar{F}_{[cd}^{\hspace{12pt}f} \sigma_{e]f}
+28 \bar{F}_{cde}\ .
\ee
To find
$D_{\bar{\alpha}} D_\beta T_{\gamma_1 a}^{\gamma_2}$,
we use the standard statement
\bb
[D_A,D_B\}=-T_{AB}^C D_C
-{\hat{R}}_{ABC}^D\ .
\ee
And we get
\bb
D_{\bar{\alpha}} D_\beta T_{\gamma_1 a}^{\gamma_2}|_0=
-T_{\bar{\alpha}\beta}^b D_b T_{\gamma_1 a}^{\gamma_2}
+R_{\bar{\alpha}\beta\gamma_1}^\delta T_{\delta a}^{\gamma_2}
+R_{\bar{\alpha}\beta a}^b T_{\gamma_1 b}^{\gamma_2}
-T_{\gamma_1 a}^{\delta} R_{\bar{\alpha}\beta\delta}^{\gamma_2}
-D_\beta D_{\bar{\alpha}} T_{\gamma_1 a}^{\gamma_2}\ .
\ee
We need
$D_{\alpha} D_{\bar{\beta}} T_{\gamma_1 a}^{\gamma_2}$, which is
\bbb
D_{\alpha} D_{\bar{\beta}} T_{\gamma_1 a}^{\gamma_2}|_0&=&
-D_\alpha D_{\gamma_1} T_{\bar{\beta} a}^{\gamma_2}
-T_{\bar{\beta}\gamma_1}^b D_{\alpha} T_{ba}^{\gamma_2}
-D_{\alpha} R_{\bar{\beta} a \gamma_1}^{\gamma_2} \nonumber \\
&-&T_{\gamma_1 a}^{\bar{\alpha}} \sigma^{b}_{\bar{\alpha}\bar{\beta}}
\sigma_{b}^{\gamma_2 \delta} D_{\alpha} D_{\delta} \omega
+T_{\gamma_1 a}^{\bar{\alpha}} \delta^{\gamma_2}_{\bar{\alpha}}
\delta_{\bar{\beta}}^\delta D_{\alpha} D_\delta \omega
+T_{\gamma_1 a}^{\bar{\alpha}} \delta_{\bar{\beta}}^{\gamma_2}
\delta_{\bar{\alpha}}^\delta D_\alpha D_\delta \omega\ .
\eee

Finally, we collect the zeroth order components of some of the superfields
that arise in the computation as well. These can be found in~\cite{HOWEWEST},
but we list them for completeness:
\bb
T_{\alpha\bar{\beta}}^a|_0=-i\sigma^a_{\alpha\beta}\ .
\ee
\bb
T_{a\beta}^{\bar{\gamma}}|_0=
-\frac{3}{16} \sigma_\beta^{bc\gamma}\bar{F}_{abc}
-\frac{1}{48} \sigma_{abcd\beta}^{\hspace{25pt}\gamma} \bar{F}^{bcd}\ .
\ee
\bb
T_{a\beta}^\gamma|_0=i \sigma^{bcde\gamma}_\beta Z_{abcde}\ .
\ee
\bb
R_{\alpha\beta,ab}|_0=i\frac{3}{4} \sigma^c_{\alpha\beta}
\bar{F}_{abc}
+\frac{i}{24} \sigma_{abcde\alpha\beta}\bar{F}^{cde}\ .
\ee
\bb
R_{\alpha\bar{\beta},ab}|_0=-\frac{1}{24}\sigma^{cde}_{\alpha\beta}
g_{abcde}\ .
\ee
\bb
H_{a\beta\gamma}|_0=-i\omega \sigma_{a\beta\gamma}\ .
\ee
\bb
H_{a\bar{\beta}\bar{\gamma}}|_0=-i\omega \sigma_{a\beta\gamma}\ .
\ee
All other components as they arise in the expansion
are zero.
The final result is given in~\pref{I4}, rescaled to the string frame as
in~\pref{I0} and~\pref{I2}.

\section{Discussion}

There is a great deal of physical information in the various parts of 
equations~\pref{I2} and~\pref{I4}. 
We first note that the terms are
linear or quadratic in the string coupling $e^\phi=\omega^2$, 
except for the canonical couplings to the Riemann tensor,
which are independent of
$\omega$. Some of the significance
of this will become clear below, as we look at the NCOS example.
The form of the action is such that the fermionic variables
$\theta$ may acquire a non-trivial vacuum configuration depending
on the strengths of the various background fields.
There is also an unusual coupling to
the derivative of the five-form field strength. And many of the
terms vanish when one considers center of mass motion of the closed
string within certain ansatzs. An important program
is then to arrange for
simplified settings and see how turning on the various couplings
independently affects the evolution of the bosonic coordinates $x^m$.
This can help us develop intuition 
about the effect of RR fields on closed string dynamics.
We defer such a  complete analysis
to an upcoming work~\cite{VVSNEW}, and content ourselves for now with a few
brief observations relevant to the NCOS case.

By substituting
into the action the fields describing the near horizon geometry 
of $(N,M)$ strings, we can study the strong coupling dynamics of
two dimensional NCOS theory. A sector of this dynamics was studied 
in~\cite{VVSlargeM}, where, by focusing on an ansatz expected to
correspond to supersymmetric trajectories, the effects of the RR
fields were ignored. For these motions, the fermions had to have
zero condensates $\lk<\theta\re>=0$. While this is reasonable to expect 
for part of the spectrum, and specially for BPS dynamics
generally associated with center of mass motion, we may find
that there are regimes or certain subsets of the solutions for which the
situation changes. Indeed, the original motivation for the current
work was to understand the extent to which this assumption is justified,
given the particulars of some of the results of~\cite{VVSlargeM}.
If there are static non-trivial condensates for the fermions,
the dynamics of the bosonic sector of the action will 
indirectly get affected.

Looking
at the form of our action as applied to the NCOS case, and looking 
for configurations near $\lk<\theta\re>=0$, we focus on the
quadratic terms given in~\pref{I2}. We can see that
only two terms contribute to the dynamics of interest; these are
of the form $\omega^2 f^{-+a}\bar{\theta}\sigma^-\theta$ and
$\omega^2 q_a \bar{\theta}\sigma^-\theta$. 
Given the form of the dilaton in these terms,
it may also be seen that the dependence on the D-string charge $M$
cancels, so that the large or infinite $M$ limit is regular on the worldsheet.
These two terms are then finite and important to the dynamics.
The first term, the one involving $f^{-+a}$, which is found
proportional to the RR 3-form once the complex conjugate piece is added, 
is necessarily multiplied by a factor $V_1^+$, which changes sign depending
on the orientation of the closed string along the direction parallel 
to the $(N,M)$ strings.
We then expect sensitivity of the effect of the RR fields on
the orientation of the strings in the solutions of~\cite{VVSlargeM};
in particular, we may hope to find that the case with 
negatively wound strings, a scenario that was already pointed out
in~\cite{VVSlargeM} to be pathological, may be consequently ruled out. 
Another very interesting effect is hidden in the second term, which
involves coupling to the axion's field strength. The axion 
has a non-trivial profile. It is attracted
to constants values inside the non-commutative throat and far away from it.
In between, a kink profile results in a flux of axion charge at the throat only!
The term $q_a \bar{\theta} \sigma^- \theta$
then becomes important at the throat.
This may be signalling that RR fields play a crucial role in 
understanding how to extend the Maldacena duality beyong the near horizon
region~\cite{VVSlargeM}. We hope to report on definitive conclusions
and a detailed analysis of all these issues in~\cite{VVSNEW}.

Other future directions include writing down the IIA action
in a similar manner, or by using T-duality (see, for example,~\cite{CVETIC}). 
Furthermore,
given the algebraic complexity of the computations involved in deriving 
some parts of our action, it can be
useful to have some of the details of our results
checked independently, preferably with different methods. 
A most ambitious, yet a very
important matter, would be to try to understand the open-closed string
duality, for example in the context of the NCOS theory, directly on the
worldsheet level. For such a map to exist, knowledge of
the couplings to the RR fields is obviously very useful. Finally, 
it would be helpful to develop general computational techniques that allow us
to analyze, at least semi-classically, dynamics of closed strings
in arbitrary backgrounds (with the effect of the RR fields we discussed
taken into account). In this regard,
approximation methods such as expansion about center of mass motion
-- which is in some respects an extension of the normal coordinate
expansion technique we used in superspace -- may be used.

\section{Appendix A: Spinors and conventions}

Our spinors are Weyl but not Majorana. They are then complex and have
sixteen components. The associated $16\times 16$ gamma matrices satisfy
\bb
\lk\{\sigma^a,\sigma^b\re\}=2\eta^{ab}\ ,
\ee
with the metric
\bb
\eta_{ab}=diag(+1,-1,-1,...,-1)\ .
\ee
Note that the signature is different from
the standard one in use in modern literature. This is so that
we conform to the equations appearing in~\cite{HOWEWEST}.
Also, the worldsheet metric $h^{ij}$ has signature $(-,+)$
for space and time, respectively.
Throughout,
the reader may refer to~\cite{HOWEWEST} to determine more about the
spinorial algebra and identities that we are using. However, 
we make no distinction between $\sigma$ and $\hat{\sigma}$ as defined
in~\cite{HOWEWEST} as this will be obvious from the context.

We note that
$\sigma^a$, $\sigma^{abcd}$ and $\sigma^{abcde}$ are symmetric; while
$\sigma^{ab}$ and $\sigma^{abc}$ are antisymmetric; and
$\sigma^{abcde}$ is self-dual.

With the choice given in~\pref{sigmapm}, we then have
\bb
\sigma^+\sigma^-+\sigma^-\sigma^+=1\ .
\ee
And complex conjugation is defined so that
\bb
\overline{\sigma^a}=\sigma^a\ .
\ee
Conjugation also implies
\bb
\overline{\theta_1 \theta_2}=\bar{\theta}_2 \bar{\theta}_1\ .
\ee
Finally, antisymmetrization is defined as
\bb
\sigma^{ab}\equiv \sigma^{[a} \sigma^{b]}\ ,
\ee
with a conventional $2!$ hidden by the braces. 

Using the completeness relation and the algebra above, we have,
for any matrix $Q_{\alpha\beta}$ with lower indices
\bb
Q_{\alpha\beta}=\frac{1}{16}\lk(
\mbox{Tr} [Q \sigma_a] \sigma^a_{\alpha\beta}
-\frac{1}{3!} \mbox{Tr} [Q\sigma_{abc}] \sigma^{abc}_{\alpha\beta}
+\frac{1}{5!} \mbox{Tr} [Q\sigma_{abcde}] \sigma^{abcde}_{\alpha\beta}
\re)\ .
\ee
This allows us, for example, to rearrange certain combinations such as
\bb
(\bar{\theta}\sigma^{-(r)}\theta)
(\bar{\theta}\sigma^{-(s)}\theta)
=\frac{1}{2} \frac{sgn(r)}{16^2}
(\bar{\theta}\sigma^-\sigma^{bc}\bar{\theta})
(\theta\sigma^-\sigma^{ef}\theta)
\mbox{Tr}
[\sigma^{bc}\sigma^{(s)}\sigma^{ef}\sigma^{(r)}]\ ,
\ee
\bb
sgn(r)\equiv \lk\{ 
\begin{array}{cc}
+1 & \mbox{for  }r=0 \\
-1 & \mbox{for  }r=2 \\
+1 & \mbox{for  }r=4
\end{array}\ ;
\re.
\ee
this identity arises repeatedly in the computations.
Finally, to avert confusion, we also note the summation convention used
\bb
U^A V_A=U^a V_a+U^\alpha V_\alpha-U^{\bar{\alpha}} V_{\bar{\alpha}}\ .
\ee

\section{Acknowledgments}
I thank P. Argyres, T. Becher, and M. Moriconi for discussions. 
I am grateful to the organizers and staff of IPAM for their warm
hospitality. This work was supported in part by a grant from the NSF.

\providecommand{\href}[2]{#2}\begingroup\raggedright\endgroup

\end{document}